\documentclass{ar-1col-S2O}

\usepackage[comma]{natbib}
\usepackage{upgreek}
\usepackage{amssymb}
\usepackage{amsmath}

\jname{Annu. Rev. Fluid Mech.}
\jvol{56}
\jyear{2024}
\doi{10.1146/annurev-fluid-121021-015818}

\newcommand\Rey{\mbox{\rm Re}}  
\newcommand\Mac{\mbox{\rm Ma}}  
\newcommand\Kn{\mbox{\rm Kn}}  

\definecolor{dgreen}{RGB}{5, 128, 20}
\definecolor{dred}{RGB}{162, 20, 47}
\definecolor{dyellow}{RGB}{200, 156, 76}

\begin{document}

\markboth{Capecelatro and Wagner}{Gas-Particle Dynamics in High-Speed Flows}

\title{Gas-Particle Dynamics in High-Speed Flows}

\author{Jesse Capecelatro$^1$ and Justin L. Wagner$^2$
\affil{$^1$Department of Mechanical Engineering and Department of Aerospace Engineering, University of Michigan, Ann Arbor, Michigan 48109; email: jcaps@umich.edu}
\affil{$^2$Engineering Sciences Center, Sandia National Laboratories, Albuquerque, New Mexico 87185;
email: jwagner@sandia.gov\\}
\affil{When citing this paper, please use the following: Capecelatro J, Wagner JL. 2024. Gas-Particle Dynamics in High-Speed Flows. Annu. Rev. Fluid Mech. 56: Submitted. DOI:10.1146/annurev-fluid-121021-015818}}

\begin{abstract}
High-speed disperse multiphase flows are present in numerous environmental and engineering applications with complex interactions between turbulence, shock waves, and particles. Compared to its incompressible counterpart, compressible two-phase flows introduce new scales of motion that challenge simulations and experiments. This review focuses on gas-particle interactions spanning subsonic to supersonic flow conditions. An overview of existing Mach number-dependent drag laws is presented, with origins from 18th-century cannon firings, and new insights from particle-resolved numerical simulations. The equations of motion and phenomenology for a single particle are first reviewed. Multi-particle systems spanning dusty gases to dense suspensions are then discussed from numerical and experimental perspectives.
\end{abstract}

\begin{keywords}
particle-laden flow, shock-particle interactions, compressible flow, drag force, turbulence, multiphase aeroacoustics
\end{keywords}
\maketitle


\section{INTRODUCTION}
High-speed (compressible) flows laden with solid particles or liquid droplets can be found across a broad range of engineering and scientific disciplines. These include naturally occurring processes, such as supernovas~\citep{inoue2009turbulence} 
and volcanic eruptions \citep{lube2020multiphase},
and human-caused flows, such as coal dust explosions \citep{griffith1978dust}; shock wave lithotripsy \citep{lingeman2009shock}; combustion/detonation \citep{zhang2001explosive}; and plume-surface interactions during space exploration \citep{capecelatro2022modeling}. While the last several decades have seen significant advancements in understanding and modeling \textit{incompressible} particle-laden flows \citep{crowe1996numerical,balachandar2010turbulent,fox2012large,tenneti2014particle,brandt2022particle}, much less attention has been paid to particle-laden \textit{compressible} flows. This article presents a review and perspectives on this topic. We focus on flows characterized by finite Mach numbers and high Reynolds numbers containing dilute to dense concentrations of rigid particles. 

Compared to incompressible flows, the examples listed above introduce physical processes taking place over a much wider range of scales. 
As an illustrative example, consider a grain of sand with diameter $d_p=100$ $\upmu$m and density $\rho_p=3000$ kg/m$^3$ in air. The particle response time due to drag is $\tau_p=\rho_p d_p^2/(18\mu_g)\approx0.1$ s, where $\mu_g$ is the gas viscosity. Meanwhile, the characteristic acoustic timescale (time for a disturbance traveling at the speed of sound $c$ to pass over the particle) is $\tau_a=d_p/c\approx0.3$ $\upmu$s, almost 6 orders of magnitude smaller! This discrepancy in time scales adds significant challenges to numerical predictions and experimental diagnostics. In addition, unsteady forces typically negligible for incompressible gas-solid flows (e.g., added mass and Basset history) can be important owing to the large acceleration difference between the phases. 
A theoretical/phenomenological description of incompressible particle-laden flows is, therefore, incomplete when gas-phase compressibility is important.

\begin{figure}[h]
\includegraphics[width=0.95\textwidth]{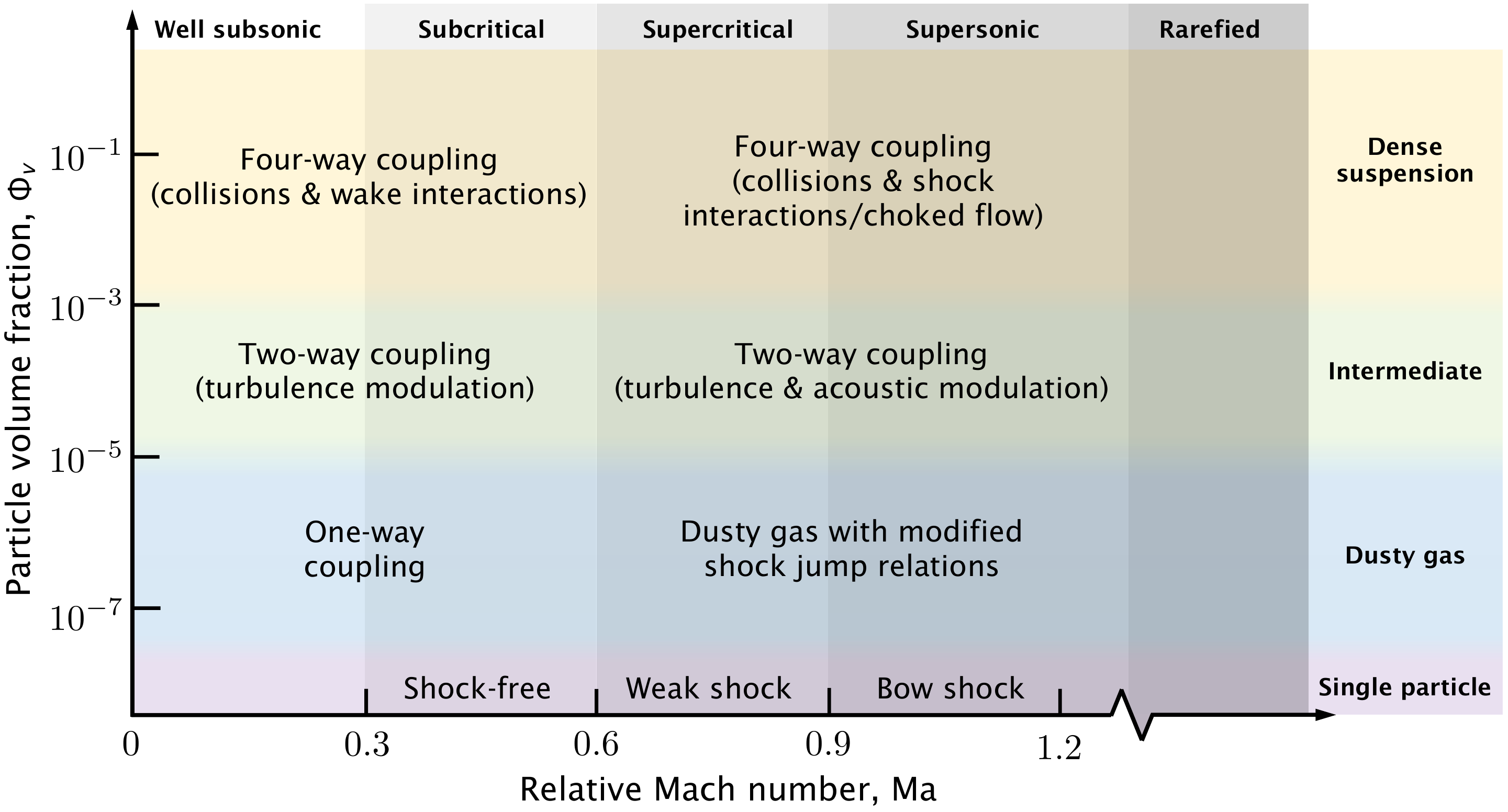}
\caption{Different regimes characterizing gas-particle interactions in high Reynolds number flows. This review covers gas-particle flows ($\rho_p/\rho_g\gg1$) in the continuum regime ($\Mac/\Rey\ll 1$). One-way coupling is applicable to incompressible and subsonic (shock-free) flows at low volume fractions. At higher Mach numbers but still low volume fractions, particles are capable of modifying shock structures. At higher volume fractions and low Mach numbers, momentum exchange between the phases is capable of enhancing or attenuating gas-phase turbulence. Dense suspensions in high Mach number flows correspond to explosive dispersal of particles with strong shock-particle-turbulence interactions.}
\label{fig:regime}
\end{figure}

An approximate regime diagram highlighting gas-particle interactions from the single particle limit to dense suspensions and Mach numbers ranging from well subsonic to supersonic is shown in \textbf{Figure~\ref{fig:regime}}. The terms subcritical and supercritical are used to denote Mach numbers below and above the value where supersonic flow around a fixed particle first occurs ($\Mac\approx 0.6$ for an isolated particle). This review focuses on gas-particle flows, where the particle-to-fluid density ratio is $\rho_p/\rho_g\gg1$, and the gas phase is in the continuum regime, characterized by Knudsen numbers $\Kn\ll 1$. For an ideal gas, the Knudsen number can be defined in terms of the particle Reynolds number and Mach number according to
\begin{equation}\label{eq:Kn}
    \Kn=\sqrt{\frac{\pi\gamma}{2}}\frac{\Mac}{\Rey},
\end{equation}
where $\gamma$ is the ratio of specific heats of the gas. The Reynolds number and Mach number here are defined in terms of the relative velocity between the gas and particle, i.e., $\Rey=\rho_g |\mathbf{u}_g-\mathbf{v}_p|d_p/\mu_g$ and $\Mac=|\mathbf{u}_g-\mathbf{v}_p|/c$, where $\mathbf{u}_g$ is the gas-phase velocity in the vicinity of the particle and $\mathbf{v}_p$ is the particle's velocity. 

\begin{marginnote}[]
\entry{CONTINUUM REGIME}{$\Kn\le 0.01$}
\entry{SLIP FLOW}{$0.01<\Kn\le0.1$}
\entry{TRANSITIONAL FLOW}{$0.1<\Kn\le10$}
\entry{FREE MOLECULAR FLOW}{$\Kn>10$}
\end{marginnote}

The flow is categorized as continuum when $\Kn<10^{-2}$. When $\Kn$ is higher, the collision rate between gas molecules and the surface of particles becomes insufficient to satisfy the no-slip condition. At $\Kn$ values between $10^{-2}$ and $10$, the flow exhibits a small departure from no-slip (slip regime). For $\Kn>10$, collisions between gas molecules and particles are frequent, while inter-molecule interactions are rare (free molecular flow). 
Thus, continuum flows at finite Mach numbers are associated with large Reynolds numbers. Two-phase rarefied flows and hypersonic flows, while rich in physics and present in many important applications, are not discussed here.


When a shock wave passes through a dusty gas, its thickness and change in pressure differ greatly from a shock  passing through an unladen gas \citep{carrier1958shock}. This is markedly different from incompressible flows, in which particles have negligible effect on the carrier phase when the particle volume fraction is $\Phi_v<10^{-5}$, termed one-way coupling \citep{elghobashi1994predicting}. Beyond the dusty gas regime, finite size particles are capable of modifying the carrier-phase turbulence (termed two-way coupling), and at moderate Mach numbers they are also capable of modifying the pressure field, resulting in acoustic or sound modulation \citep{crighton1969sound,krothapalli2003turbulence,buchta2019sound}. At even higher volume fractions, collisions between particles and fluid-mediated wakes drive particle dynamics (termed four-way coupling). When the flow is compressible at these volume fractions, the interstitial space between particles creates a nozzling effect that results in choked flow \citep{theofanous2018shock}.

This review provides an up-to-date account on the current understanding and modeling capabilities of high-speed particle-laden flows and concludes with a brief perspective on future research directions.  We begin with the equations of motion and phenomenology for a single particle. We then focus on multi-particle systems spanning dusty gases to dense suspensions from both a numerical and experimental perspective, including multiphase aeroacoustics, turbulence induced during shock-particle interactions, and drag at finite volume fraction and Mach number.

\section{PARTICLE EQUATION OF MOTION}
The fluid force acting on a particle can be decomposed into separate contributions: 
the quasi-steady drag force $\mathbf{F}_{qs}$, undisturbed flow forces $\mathbf{F}_{un}$ (sometimes denoted the pressure gradient or Archimedes force), inviscid unsteady force $\mathbf{F}_{iu}$ (often referred to as added-mass, see the sidebar titled Added Mass in a Compressible Fluid), and viscous-unsteady force $\mathbf{F}_{vu}$ (Basset history), together expressed as
\begin{equation}\label{eq:BBO}
    m_p\frac{{\rm d}\mathbf{v}_p}{{\rm d}t}=\mathbf{F}_{qs}+\mathbf{F}_{un}+\mathbf{F}_{iu}+\mathbf{F}_{vu},
\end{equation}
where $m_p$ is the particle mass. \citet{maxey1983equation} and \citet{gatignol1983faxen} derived expressions for each term in the context of a spherical particle moving through an incompressible fluid at low Reynolds numbers, which are widely employed in numerical simulations.

\begin{textbox}[h]\section{ADDED MASS IN A COMPRESSIBLE FLUID}
The term added (or virtual) mass refers to the enhanced inertia of an object caused by the surrounding volume of fluid moving with it. This yields an inviscid unsteady force that can be expressed in terms of a response kernel, $K_{iu}(\tau;\Mac)$, used to weigh the history of the particle's acceleration, as shown in Equation~\ref{eq:Fiu}. In an incompressible flow, sound propagates infinitely fast and the kernel reduces to a Dirac delta function, $K_{iu}(\tau;\Mac=0)=\delta(\tau)/2$, allowing it to be written as the product of the relative acceleration and an added-mass coefficient ($C_m$). For a spherical particle in an incompressible flow, integration of the kernel yields $C_m=0.5$. In compressible flow, the kernel decays over a short but finite time that depends on the particle's shape and Mach number. Consequently, the force no longer takes the form of a constant mass multiplied by the instantaneous acceleration. Because of this, \citet{miles1951virtual} and \citet{longhorn1952unsteady} emphasized that reference to this force as a `virtual' or `added' mass is only applicable to incompressible flows.
\end{textbox}


For viscous compressible flows, the separate force contributions are given by \citep{parmar2011generalized,parmar2012equation}
\begin{eqnarray}\label{eq:forces}
\mathbf{F}_{qs}  = \frac{1}{2} C_D\rho_g\left(\mathbf{u}_g-\mathbf{v}_p\right)|\mathbf{u}_g-\mathbf{v}_p|A_p,\label{eq:Fqs} \\
\mathbf{F}_{un}  = V_p\rho_g\frac{{\rm D}\mathbf{u}_g}{{\rm D}t}, \\
\mathbf{F}_{iu}  = V_p\int_{-\infty}^t K_{iu}\left(t-\chi;\Mac\right)\left(\frac{{\rm D}(\rho_g\mathbf{u}_g)}{{\rm D}t}-\frac{{\rm d}(\rho_g\mathbf{v}_p)}{{\rm d}t}\right)_{t=\chi}{\rm d}\chi,\label{eq:Fiu} \\
\mathbf{F}_{vu}  = \frac{3}{2}d_p^2\sqrt{\pi\rho_g\mu_g}\int_{-\infty}^t K_{vu}\left(t-\chi;\Rey,\Mac\right)\left(\frac{{\rm D}(\rho_g\mathbf{u}_g)}{{\rm D}t}-\frac{{\rm d}(\rho_g\mathbf{v}_p)}{{\rm d}t}\right)_{t=\chi}{\rm d}\chi,\label{eq:Fvu}
\end{eqnarray}
where $A_p$ is the frontal area of the particle, $V_p$ is its volume, and $K_{iu}$ and $K_{vu}$ are the inviscid and viscous-unsteady force kernels, respectively. Although Equations~\ref{eq:Fqs}--\ref{eq:Fvu} were derived for a single particle in the limit $\Rey\rightarrow0$ and $\Mac\rightarrow0$, they provide a framework for empirical extensions to more complex flow conditions. For example, $\mathbf{F}_{qs}$ includes a drag coefficient $C_D$ that in general depends upon $\Rey$, $\Mac$, and $\Phi_v$. Models for these expressions are typically developed using far-field flow quantities measured at a distant region away from the particle. Reconstructing far-field quantities at the particle location can be challenging in numerical simulations when the particle disturbs the flow field \citep{horwitz2016accurate} or in flows involving shocks, for which these quantities may be discontinuous~\citep{jacobs2009high}. Even with accurate estimates of the far-field quantities, models valid for finite Mach numbers \textit{and} volume fractions are only starting to become available. In the following sections we summarize existing models for quasi-steady drag, the inviscid unsteady force, their origins, and extensions to multi-particle systems.

\section{SINGLE PARTICLE FLOW}
\begin{figure}[h]
\includegraphics[width=1\textwidth]{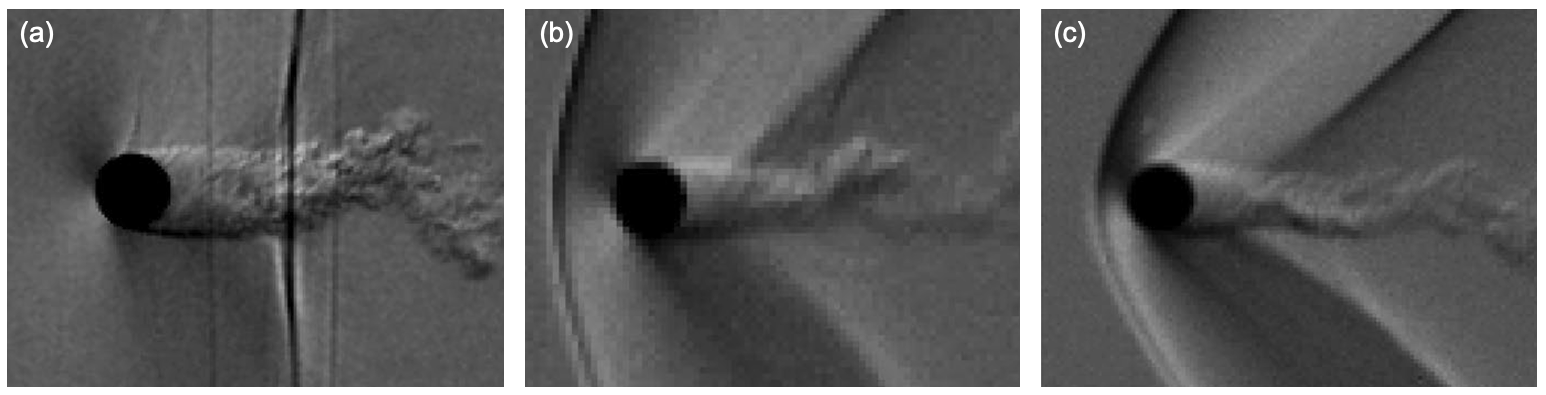}
\caption{Schlieren visualization of flow past a sphere with (\textit{a}) $\Mac=0.9$, (\textit{b}) $\Mac=1.21$, and (\textit{c}) $\Mac=1.39$. Adapted from \citet{nagata2020experimental} with permission.}
\label{fig:Nagata}
\end{figure}

In this section, we review the state-of-the-art in modeling the forces acting on an \textit{isolated} particle. 
Such an example is provided in \textbf{Figure~\ref{fig:Nagata}}. We then move on to extensions to multi-particle systems.

\subsection{Quasi-Steady Drag ($\mathbf{F}_{qs}$)}
\begin{figure}[h]
\includegraphics[width=0.9\textwidth]{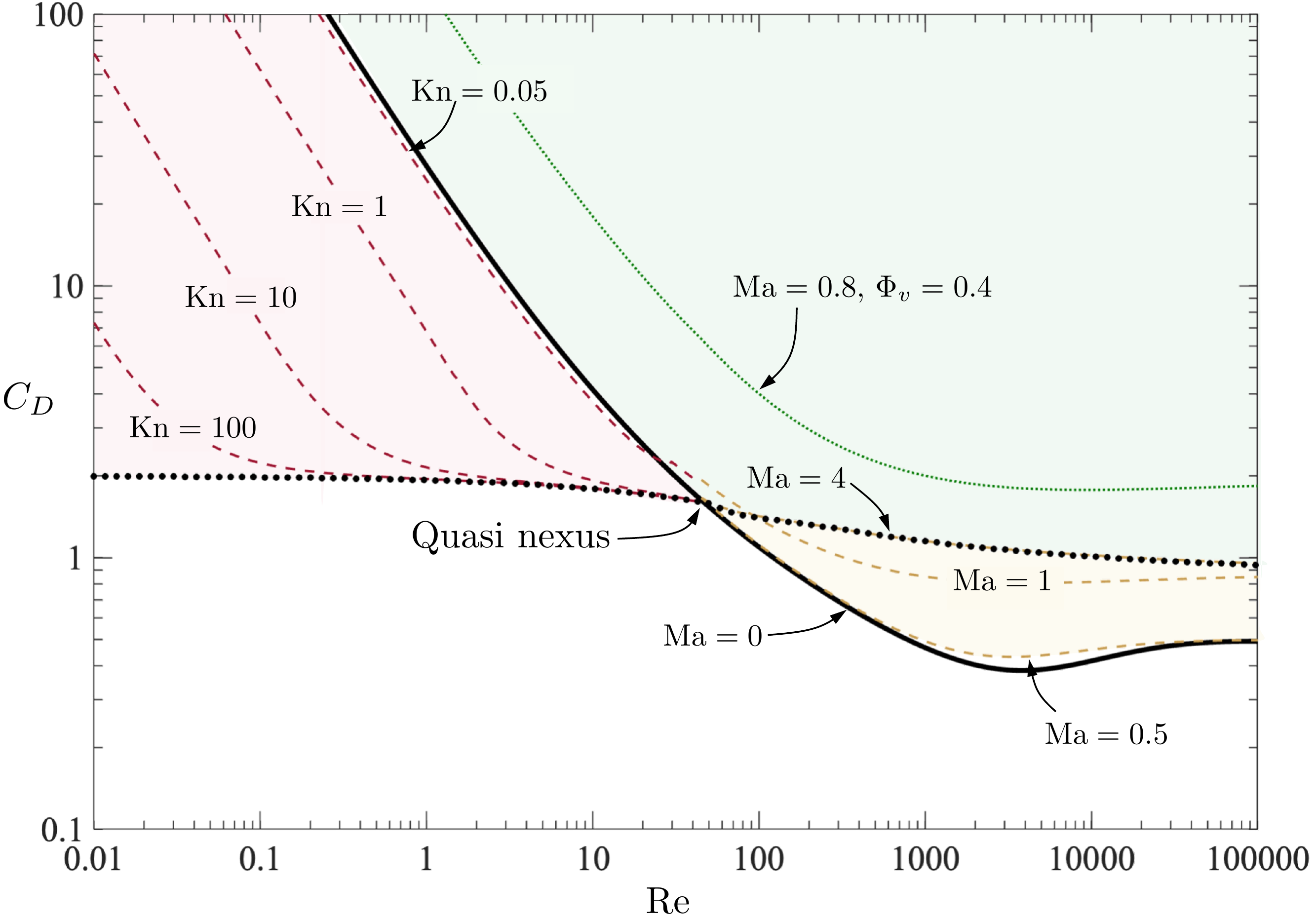}
\caption{Drag coefficient of a sphere highlighting the rarefaction dominated regime (red), compression dominated regime (yellow), and multi-particle continuum regime (green). Incompressible flow ($\Mac=0$) past an isolated particle \citep{clift1971motion} ({\bfseries --}). Free-molecular flow ($\Mac\gg 0$ and $\Kn\gg 0$) past an isolated particle ({\tiny $\bullet\,\bullet\,\bullet$}). Single-particle correlation of \citet{loth2021supersonic} (dashed lines). 
Multi-particle correlation of \citet{Osenes2023drag} (\textcolor{dgreen}{$\cdots$}). Adapted from \cite{loth2021supersonic} with permission.}
\label{fig:drag}
\end{figure}

A culmination of experimental and semiempirical studies from the 20th century yielded reliable estimates of $C_D$ for a sphere in incompressible flows up to $\Rey=10^5$~\citep[e.g.,][]{oseen1910uber,goldstein1929steady,schiller1933fundamental,clift1971motion}. General trends of the drag coefficient as a function of Reynolds number are shown in \textbf{Figure~\ref{fig:drag}}. The fit by \citet{clift1971motion} is among the most comprehensive for incompressible flows. At higher speeds, the drag force is complicated by the emergence of expansion fans and shock waves. For subcritical Mach numbers ($\Mac \lessapprox 0.6$), the drag coefficient is only weakly affected by compressibility due to the absence of shocks. For supercritical but still subsonic Mach numbers, $C_D$ increases sharply with Mach number due enhanced pressure by a weak shock. At supersonic speeds, a bow shock is formed--with a stand-off distance that decreases with increasing Mach number--that leads to a large increase in $C_D$.
Schlieren imaging of a sphere in free flight are shown in \textbf{Figure~\ref{fig:Nagata}}, highlighting these different regimes.

\begin{textbox}[h]\section{ORIGINS OF MODERN-DAY DRAG LAWS}
Book II of Isaac Newton's \textit{Principia} was one of the earliest works to estimate the drag of a sphere, demonstrating the force is proportional to the square of the object's speed through the fluid $U$, its cross-sectional area $A$, and the density of the carrier fluid $\rho$: $F_d=\rho A U^2C_D/2$. His early experiments showed the drag coefficient of a sphere to be $C_D\approx0.5$ at low speeds. Half a century later, British mathematician Benjamin Robins provided the first measurements of drag on a sphere traveling at high speeds following his invention of the ballistic pendulum. Using round shots fired from guns, drag was found to scale as $U^3$, with values of $C_D$ as much as three times greater than Newton's original estimate \citep{howard1742new}.

Further progress was not made until more than a century later when Francis Bashforth conducted experiments of artillery round shots (cannon fire) for the British Army. (Bashforth with his former college classmate John Couch Adams later went on to develop the Adams--Bashforth method~\citep{bashforth1883attempt}, a class of multi-step methods commonly used for numerical time integration.) Bashforth's invention of the ballistic chronograph in 1864 allowed for up to 10 velocity measurements per shot, providing reliable estimates of acceleration over a wide range of speeds~\citep{bashforth1870}. Cannon firings operate in a fortuitous flow regime for studying drag. Typical velocities span $100-700$ m/s, corresponding to Mach numbers ranging from 0.3 to 2 in air, where $C_D$ changes sharply due to compressibility effects. Further, diameters range from $50-200$ mm, corresponding to Reynolds numbers near the critical value ($\Rey\approx2\times10^5$). The drag force was found to scale according to $F_d\propto U^2$ at moderate (subsonic) velocities, consistent with Newton. At higher velocities (subsonic to moderately supersonic) he showed $F_d\propto U^3$, consistent with Robins, and at even higher velocities it is again proportional to $U^2$~\citep{gilman1905ballistic}. 

More than a century later, \citet{miller1979sphere} compiled available data for drag on a sphere at moderate to high Mach numbers and found Bashforth's measurements to be among the most accurate. This data, combined with more recent free-flight measurements from aeroballistic ranges, continues to be used in the development of modern-day drag laws \citep{clift2005bubbles}. 
\end{textbox}


One of the earliest Mach number-dependent drag laws was developed using free-flight measurements from aeroballistic ranges \citep{henderson1976drag}. It includes non-continuum effects when the mean-free path of the gas phase approaches the particle diameter (i.e., $\Kn>0$).   \citet{loth2008compressibility} later developed a drag coefficient by separating the flow into a rarefaction-dominated regime for $\Rey<45$ (shaded red in \textbf{Figure~\ref{fig:drag}}) and a compression-dominated regime when $\Rey>45$ (shaded yellow in \textbf{Figure~\ref{fig:drag}}). In between, it was suggested that $C_D\approx1.63$ is independent of $\Mac$ and $\Kn$. 
However, both models yield significant errors near the transonic regime \citep{parmar2010improved}.
As described in \citet{clift2005bubbles}, much of the data used for constructing Mach number dependent drag laws are unreliable due to high levels of freestream turbulence, interference by supports, and wall effects. It is interesting to note that many of these correlations are formulated from data that can be traced back to the 18th century (see the sidebar titled Origins of Modern-Day Drag Laws).


With the advent of high-performance computing, numerical simulations with grid spacing $\Delta x \ll d_p$, termed particle-resolved direct numerical simulations (PR-DNS) \citep{tenneti2014particle}, are beginning to shed new light on this topic.  \citet{loth2021supersonic} combined PR-DNS of \citet{nagata2020direct} with rarefied-gas simulations and an expanded experimental dataset to refine $C_D$, showing improved accuracy over existing models. 
The resulting drag coefficient is shown in \textbf{Figure~\ref{fig:drag}}. It can be seen that the compression-dominated region ($\Rey>60$) yields an increase in $C_D$ as $\Mac$ increases, whereas in the rarefaction-dominated region ($\Rey<30$), $C_D$ is inversely proportional to $\Mac$. 
Although significant progress has been made in recent years, additional data is needed for refinement and validation, particularly within the quasi-nexus region \citep{loth2021supersonic}. 

\subsection{Unsteady Forces ($\mathbf{F}_{un}$, $\mathbf{F}_{iu}$, $\mathbf{F}_{vu}$)}\label{sec:unsteady}
\begin{figure}[h]
\includegraphics[width=1\textwidth]{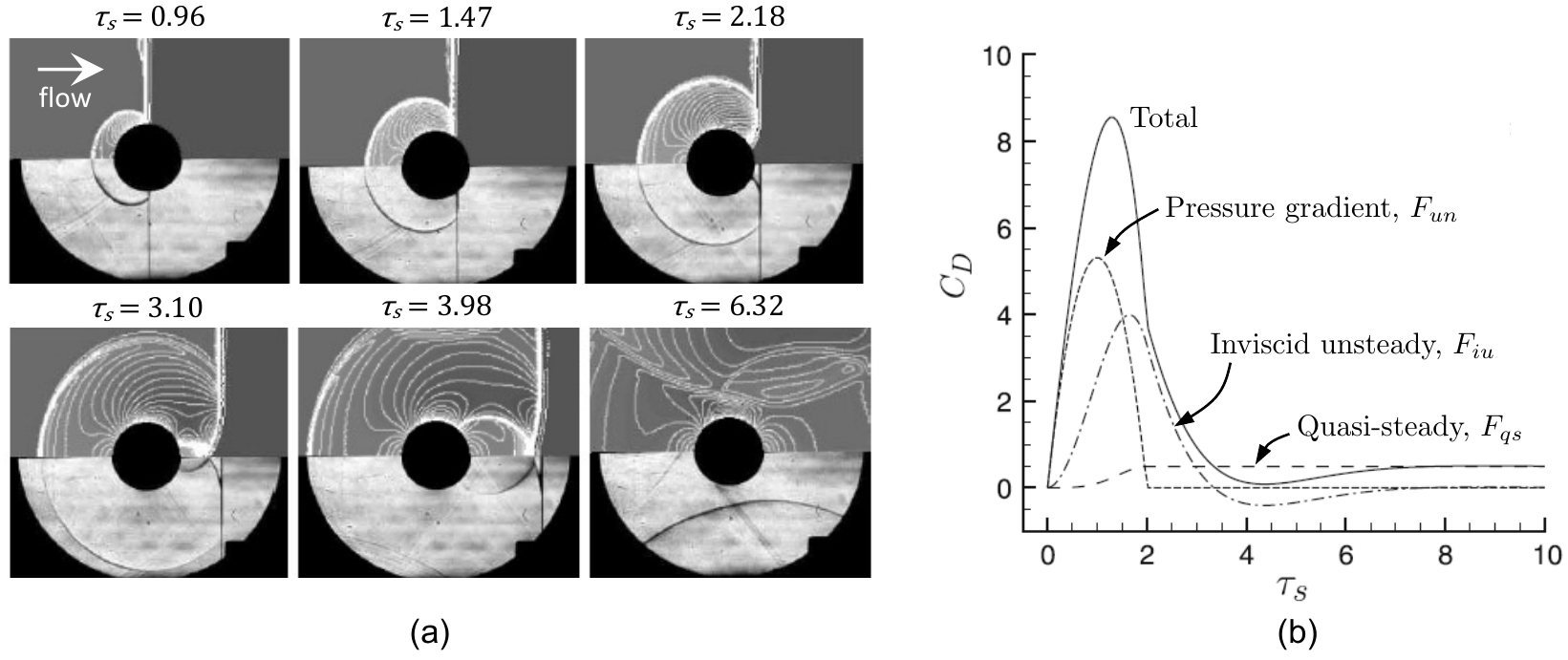}
\caption{Interaction of a normal shock with a sphere. (\textit{a}) Shock tube experiment at a shock Mach number $M_s=1.22$. (\textit{b}) Unsteady drag prediction from a numerical model. Panel \textit{a} adapted from \citet{tanno2003interaction}. Panel \textit{b} adapted from \citet{parmar2009modeling}.}
\label{fig:Tanno}
\end{figure}

Experiments involving the passage of a shock wave over a sphere have provided measurements of unsteady drag coefficients in the presence of strong acceleration \citep{britan1995acceleration,tanno2003interaction,sun2005unsteady,bredin2007drag,skews2007drag}. \textbf{Figure~\ref{fig:Tanno}\textit{a}} depicts a planar shock wave interacting with a stationary sphere in a shock tube as a function of non-dimensional time $\tau_s = 2tu_s/d_p$, where $u_s$ is the shock speed. Experimental observations and numerical simulations have shown that the peak drag coefficient occurs just before the shock reaches the sphere equator with values as much as one order of magnitude larger than the steady counterpart \citep{sun2005unsteady,bredin2007drag,skews2007drag}. Shortly after ($\tau_s\approx 2$), the shock wave is diffracted on the downstream side of the particle, resulting in the generation of high pressure that temporarily reduces drag. For weak shocks with subcritical post-shock Mach numbers ($\Mac<0.6$), this can lead to a period of negative $C_D$ \citep{osnes2022}. At sufficiently low Reynolds numbers, the negative contribution from pressure drag is counteracted by viscous forces, preventing $C_D$ from becoming negative \citep{sun2005unsteady}. The high pressure region behind the particle then expands and develops into a wake, at which point the flow transitions to a quasi-steady flow state. Note that the relative contributions of steady and unsteady forces obtained from experiments of shock-particle interactions have been debated (see the sidebar titled Perceived Unsteady Effects from Shock Tube Experiments).

\begin{textbox}[h]\section{PERCEIVED UNSTEADY EFFECTS FROM SHOCK TUBE EXPERIMENTS}
Several pioneering multiphase shock tube experiments \citep{igra1993shock,suzuki2005shock,jourdan2007drag} have reported elevated drag coefficients for small spheres (approximately 1 mm) in high $\rho_p/\rho_g$ flows compared to the `standard' incompressible drag law of \cite{clift1971motion}. These studies attributed the increase in drag to unsteady acceleration of the spheres. According to \cite{parmar2009modeling}, however, such unsteady effects should dissipate rapidly for small spheres and therefore have negligible effect on long time drag measurements. Drag coefficient data at transonic speeds in more recent shock tube experiments \citep{wagner2012shock} and a comparison to compressible drag models of \cite{loth2008compressibility} and \cite{parmar2010improved} suggested the elevated drag coefficients in previous shock tube studies are likely related to pronounced compressibility effects not captured by earlier models such as that of \cite{henderson1976drag}. These same experiments also noted increased drag even for a few widely spaced particles near transonic Mach numbers, which might also explain elevated drag measured in previous shock tube studies. Additionally, careful characterization of particle size is critical \citep{bordoloi2017relaxation, maxon2021high}.  
\end{textbox}

Unsteady contributions to the particle equation of motion are often neglected when the particle-to-fluid density ratio ($\rho_p/\rho_g$) is large. As shown in \textbf{Table \ref{table:scaling}}, the ratio of the inviscid unsteady forces ($\mathbf{F}_{iu}$ and $\mathbf{F}_{un}$) to the quasi-steady drag $\mathbf{F}_{qs}$ in Equation~\ref{eq:BBO} scales like $(\rho_p/\rho_g+C_m)^{-1}$ when a particle accelerates in a quiescent fluid, with $(\rho_p/\rho_g+C_m)^{-1/2}$  scaling for the viscous unsteady force, $\mathbf{F}_{vu}$. Thus, for gases comprised of liquid droplets or solid particles, these contributions are indeed small. In contrast, the relative contributions of the unsteady forces are \textit{independent} of density ratio when particles are placed in non-uniform flows \citep{taylor1928forces,magnaudet1995accelerated}. 
As summarized in \textbf{Table \ref{table:scaling}}, the order of magnitude of the inviscid unsteady forces compared to quasi-steady drag is $\Rey d_p/\mathcal{L}$, where $\mathcal{L}$ is a characteristic length scale of the flow. Similarly, $|\mathbf{F}_{vu}|/|\mathbf{F}_{qs}|\propto\sqrt{\Rey d_p/\mathcal{L}}$. Thus, unsteady forces arising due to fluid acceleration are independent of density ratio when $\Rey d_p/\mathcal{L}\ge\mathcal{O}(1)$. From this scaling it is apparent that unsteady forces are important during shock-particle interactions, where $\Rey\gg 1$ and the characteristic length scale corresponds to the shock thickness ($\mathcal{L}\ll d_p$), as seen in \textbf{Figure~\ref{fig:Tanno}\textit{b}}.

\begin{table}[h]
\tabcolsep7.5pt
\caption{Relative importance of the inviscid and viscous unsteady forces compared to quasi-steady drag. Adapted from \cite{bagchi2002steady}.}
\label{table:scaling}
\begin{center}
\begin{tabular}{@{}r c@{} c@{}}
\hline
& \textbf{Inviscid unsteady}, $\textbf{F}_{iu}$~~~~ & \textbf{Viscous unsteady}, $\textbf{F}_{vu}$\\
\textbf{Accelerating fluid:} & $\Rey\frac{d_p}{\mathcal{L}}$ & $\sqrt{\Rey\frac{d_p}{\mathcal{L}}}$\\
\textbf{Accelerating particle:} & $\frac{1}{\rho_p/\rho_g+C_m}$ & $\frac{1}{\sqrt{\rho_p/\rho_g+C_m}}$\\
\hline
\end{tabular}
\end{center}
\end{table}

As shown in Equation~\ref{eq:Fiu}, the inviscid unsteady force involves a kernel $K_{iu}$ that depends on prior history and local Mach number. In the limit of a sphere in incompressible flow, the kernel has a closed-form expression given by $K_{iu}(\tau;\Mac=0)=\exp(-\tau)\cos(\tau)$ \citep{longhorn1952unsteady}. Using the compressible form of the Bernoulli equation, a Mach number expansion shows the added-mass coefficient scales like $C_m\propto1+\Mac^2+\mathcal{O}\left(\Mac^4\right)$ for $\Mac<0.6$  \citep{parmar2008unsteady}. At a Mach number of $0.5$, the added-mass coefficient is $C_m(\Mac=0.5)\approx 1$, approximately twice as large as the value for a sphere in an incompressible flow.

\cite{parmar2009modeling} derived a model capable of capturing unsteady shock loading applicable to subcritical cases where the post-shock Mach number is less than 0.6 (see \textbf{Figure~\ref{fig:Tanno}\textit{b}}). The model includes the pressure gradient force ($\mathbf{F}_{un}$) and history term ($\mathbf{F}_{iu}$) while neglecting $\mathbf{F}_{vu}$, which was deemed inconsequential. Like previous experimental data and numerical simulations, the model shows the peak drag coefficient of an isolated particle interacting with a shock wave can be an order of magnitude larger than the value from the quasi-steady drag force for a wide range of Mach numbers and Reynolds numbers. These unsteady effects are active when the shock passes over the particle, but have little contribution to long term particle motion if $\rho_p/\rho_g$ is large. Although subcritical studies have been enlightening, additional work is required to understand the effects at higher Mach numbers where inviscid unsteady forces may be difficult to discern from vortex shedding \citep{parmar2009modeling}. 

The modeling efforts discussed heretofore consider flows interacting with an isolated particle. The flow through assemblies of particles, and corresponding forces that arise, can deviate significantly from the situation of a single particle. At present, accurate models validated at finite values of Reynolds number, Mach number, and volume fraction are lacking. The following sections provide an overview of numerical and experimental campaigns focused on compressible flows in multi-particle systems.

\section{EQUATIONS OF MOTION FOR MULTI-PARTICLE SYSTEMS} 
\begin{figure}[h]
\includegraphics[width=1.0\textwidth]{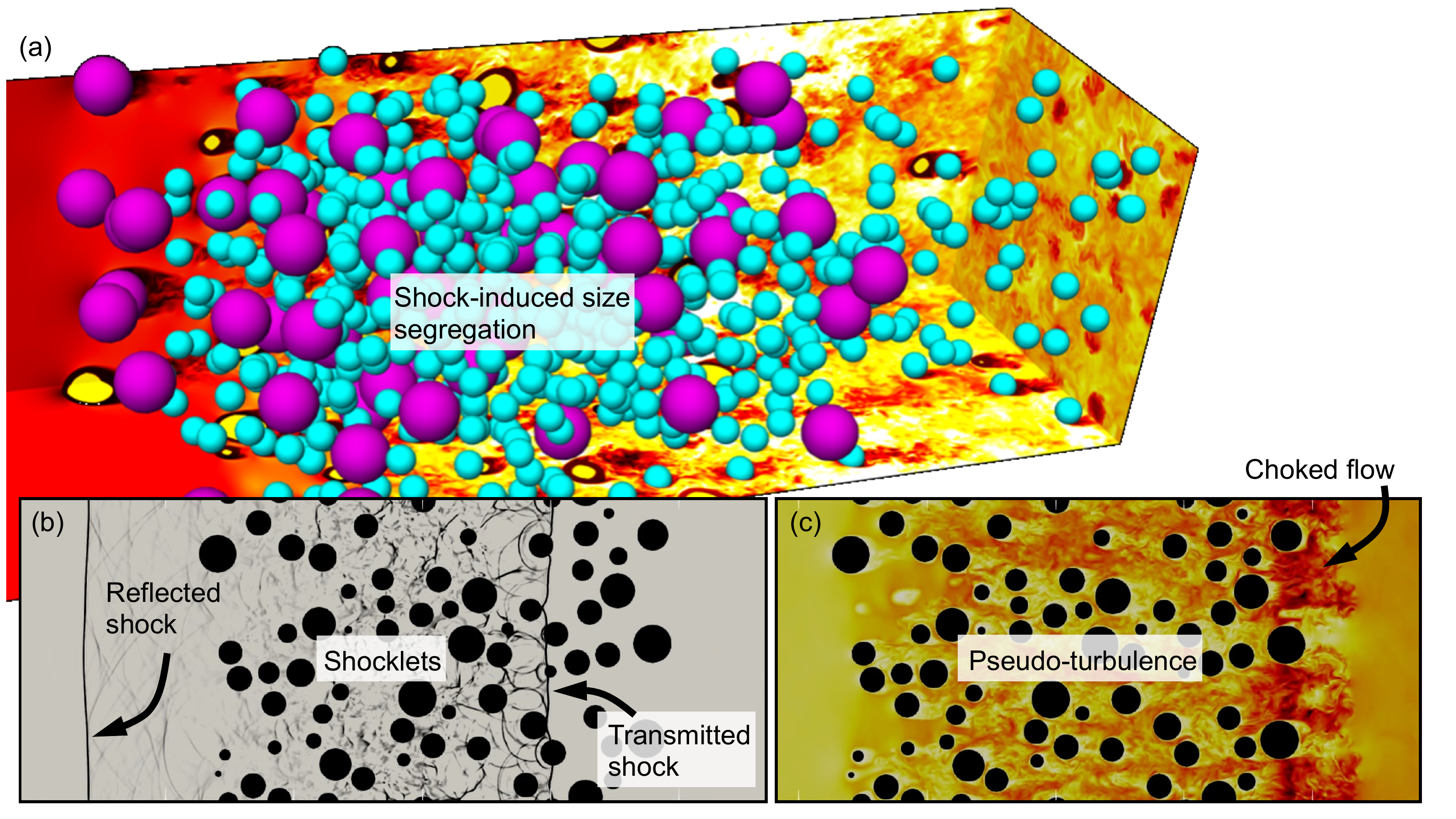}
\caption{PR-DNS of a $\Mac_s=1.66$ shock interacting with a cloud of particles with an initial volume fraction $\Phi_v=0.21$. (\textit{a}) Bidisperse distribution of particles after the shock traverses the cloud. (\textit{b}) Numerical schlieren at an early time when the shock is still within the cloud. (\textit{c}) Contour of local Mach number shortly after the shock passes through the cloud. The simulation was performed using the numerical framework outlined in \citet{khalloufi2023drag}.}
\label{fig:shock-particle}
\end{figure}

Particle-particle interactions can fundamentally alter the evolution of the two-phase flow at sufficiently high volume fractions. In incompressible flow, when $\Phi_v\gtrapprox10^{-3}$, interparticle collisions and wakes from neighboring particles directly modify the drag force and carrier-phase turbulence. In high-speed flows, the emergence of bow shocks around individual particles and reflected shocks from neighboring particles complicates this picture. An example is shown in 
\textbf{Figure~\ref{fig:shock-particle}}, where PR-DNS of a planar shock interacting with a cloud of particles reveals shocklets and pseudo-turbulence. Even shortly after the shock passes over the curtain, appreciable size segregation can be observed, with smaller particles traveling further downstream. This is consistent with the particle diameter scaling of the unsteady forces given in \textbf{Table \ref{table:scaling}}.

\begin{marginnote}[]
\entry{INTRINSIC TURBULENCE}{Turbulence that manifests from large-scale motions via an energy cascade that would exist even in the absence of particles}
\entry{PSEUDO TURBULENCE}{Turbulence originating at the scale of individual particles due to wakes and shocks}
\end{marginnote}


The mass loading, defined by the ratio of the specific masses of the
particle and fluid phases, $\Phi_m=\rho_p\Phi_v/[\rho_g(1-\Phi_v)]$, characterizes the extent to which interphase coupling is important. When $\Phi_m\ll1$, the effect of particles on the background flow is negligible. As $\Phi_m$ increases, mass, momentum, and heat transfer from the particles to the fluid become increasingly more important. In this section, we first consider the equations governing flows made up of small, non-interacting particles in the limit $\Phi_v\rightarrow 0$ and $\Phi_m>0$ (dusty gas regime) then discuss the equations describing flows containing finite size particles. 

\subsection{Dusty Gas Regime}
The dusty gas approach assumes the carrier fluid contains many small non-interacting particles ($\Phi_v\approx 0$), but the density ratio is sufficiently high such that $\Phi_m>0$. If the timescales associated with interphase exchange are small compared to the characteristic time of the flow, then equilibrium of temperature and velocity can be assumed between the two phases. Under these assumptions, the mixture density is $\rho_g(1+\Phi_m)$ and the ratio of specific heats in the mixture (denoted with an asterisk) is given by \citep{marble1970dynamics}
\begin{equation}
    \frac{\gamma^*}{\gamma}=\frac{C_p+\Phi_mC_{p,p}}{C_p+\gamma\Phi_mC_{p,p}},
\end{equation}
where $C_p$ and $C_{p,p}$ are the ratio of specific heats of the gas and particles, respectively. Consequently, the sound speed in the gas-particle mixture is
\begin{equation}
    \frac{c^*}{c}=\sqrt{\frac{\gamma^*}{\gamma}\frac{1}{1+\Phi_m}}.
\end{equation}
These relations are often employed when modeling large-scale geophysical phenomena, such as pyroclastic density currents \citep{sulpizio2014pyroclastic} and volcanic eruptions \citep{carcano2014influence,valentine2018compressible}, and shock waves in the interstellar medium \citep{draine1993theory}.


A defining feature of dusty gases is that they modify shock structures in the carrier phase, despite the low volume fraction. 
When a shock wave propagates through a dusty gas, particles remove momentum and energy from the gas, causing the strength of the shock to decay faster than it would otherwise. At sufficiently high mass loading, the shock decays to a weak pressure wave and eventually becomes fully dispersed \citep{miura1982dusty}. 
However, particles of finite size delay the time it takes the two phases to reach an equilibrium. Non-negligible slip velocities between the phases give rise to turbulence modulation (for sufficiently viscous flows) and the emergence of bow shocks (for sufficiently high Mach numbers), as depicted in \textbf{Figure~\ref{fig:shock-particle}}. Thus, for many applications, the dusty gas approximation is not valid and instead transport equations need to be solved for each phase separately.

\subsection{Volume-Averaged Equations}
In the dusty gas approach, only the momentum and energy equations of the gas phase are employed, along with a transport equation for the particle density. Extensions to two-phase flows at moderate and high volume fractions require some type of averaging of the governing equations. \citet{anderson1967fluid} applied a spatial filter to the Navier--Stokes equations to arrive at a set of equations for each phase that can be solved at a scale larger than the size of the  particle. This is analogous to large-eddy simulation (LES) of single-phase flows, and similarly results in unclosed terms that require models. Unlike in single-phase LES, the filtering procedure omits the volume occupied by particles, resulting in sub-filtered (or subgrid-scale) contributions that account for the presence of particles.

The formulation of \citet{anderson1967fluid} was recently extended to compressible flows \citep{shallcross2020volume}, which results in additional unclosed terms that need to be modeled. Volume filtering the viscous compressible Navier--Stokes equations yields
\begin{equation}\label{eq:density}
\frac{\partial (1-\Phi_v)\rho_g}{\partial t}+\nabla\mathbf{\mathbf{\cdot}}\left[(1-\Phi_v)\rho_g \textbf{u}_g\right]=0,
\end{equation}
\begin{equation}\label{eq:momentum}
\frac{\partial (1-\Phi_v) \rho_g \textbf{u}_g}{\partial t} + \nabla \mathbf{\cdot}\left[(1-\Phi_v) \left(\rho_g \textbf{u}_g\textbf{u}_g + \textbf{R}_g\right)\right]  =(1-\Phi_v)\nabla\mathbf{\cdot} \left(\mathbf{\sigma}_g-p_g\textbf{I}\right)+\textbf{F}_p,
\end{equation}
and
\begin{equation}\label{eq:energy}
\begin{split}
\frac{\partial (1-\Phi_v) \rho_g E_g}{\partial t} & + \nabla \mathbf{\cdot} \left[ (1-\Phi_v)\rho_g\textbf{u}_g(E_g+p_g)  \right] + \nabla \mathbf{\cdot} \left[(1-\Phi_v)\textbf{u}_g\mathbf{\cdot} (\textbf{R}_g -\mathbf{\sigma}_g)\right]\\
& = -(1-\Phi_v) \nabla \mathbf{\cdot} \textbf{q}_g+p_g\frac{\partial\Phi_v}{\partial t}+\mathbf{\sigma}_g\mathbf{:} \nabla \left( \Phi_v \textbf{u}_p \right)  + \textbf{u}_p \mathbf{\cdot} \textbf{F}_p + Q_p,
\end{split}
\end{equation}
where $E_g$ the total gas-phase energy, $p_g$ is the gas-phase pressure, $\mathbf{\sigma}_g$ is the viscous stress tensor, $\textbf{I}$ is the identity matrix, $\textbf{q}_g$ is the gas-phase heat flux, and $\textbf{u}_p$ is the particle-phase velocity in an Eulerian frame of reference. Momentum exchange from the particles to the gas are accounted for in $\textbf{F}_p$, which contains the fluid contributions on the right-hand side of Equation~\ref{eq:BBO}. Heat exchange from the particles to the gas is captured in $Q_p$, which is typically modeled using a Nusselt-number correlation \citep{ling2016inter,das2018metamodels}. 

Because carrier-phase velocity fluctuations may originate at the particle scale, the residual stress $\textbf{R}_g$ differs significantly from the Reynolds stress appearing in classical single-phase turbulence. This term may even be non-zero in laminar flows due to unresolved particle wakes and is therefore termed a \textit{pseudo} turbulent Reynolds stress~\citep{mehrabadi2015pseudo}. Recent work has shown that the pseudo-turbulent kinetic energy (PTKE), defined as $k_g={\rm tr}(\textbf{R}_g)/2$, can contribute to a significant portion of the total kinetic energy during shock-particle interactions \citep{hosseinzadeh2018investigation,sen2018role,mehta2019Pseudo,osnes2019computational,shallcross2020volume} and plays an important role in satisfying conservation \citep{fox2020hyperbolic}. Algebraic models for PTKE in incompressible \citep{mehrabadi2015pseudo} and compressible \citep{osnes2019computational} flows, and transport equations for compressible flows \citep{shallcross2020volume} have been proposed, but such models are still in their infancy. Future modeling efforts must take care when distinguishing between velocity fluctuations originating from turbulent motion generated at scales larger than particles (intrinsic turbulence) and those induced by particles (pseudo turbulence).

\subsection{Ill-Posedness}
The averaging procedure discussed above can be applied to both phases, resulting in Eulerian-based two-fluid models \citep[e.g.,][]{houim2016multiphase}, or combined with a Lagrangian description of the particle phase, the so-called Eulerian--Lagrangian approach \citep{patankar2001modeling,capecelatro2013euler}. An important requirement for a well-defined two-fluid model is that its closure models must ensure that the system of equations is hyperbolic. Otherwise, their solutions may yield complex eigenvalues. This is generally not an issue in Eulerian--Lagrangian methods due to the Lagrangian treatment of the particles. However, the compressible two-fluid equations are known to become ill-posed due to lack of hyperbolicity when two-way coupling is accounted for, namely through the Archimedes force acting on the particles ($\mathbf{F}_{un}$ in Equation~\ref{eq:BBO}) \citep{lhuillier2013quest}. This has been shown to generate spurious volume fraction disturbances in simulations of shock-particle interactions \citep{theofanous2018shock} and degenerate solutions in simulations of supersonic jet-induced cratering \citep{balakrishnan2021fluid}.

Since the 1970s, numerous attempts have been made to address the lack of hyperbolicity, namely by adding ad-hoc forces to stabilize the solution \citep{stuhmiller1977influence}. This was recently resolved by \citet{fox2019kinetic}, who derived a hyperbolic two-fluid model starting from the Boltzmann--Enskog kinetic equations, wherein the fluxes and source terms have unambiguous definitions. The following year, \citet{fox2020hyperbolic} extended the model to arbitrary density ratios by including the added mass of the fluid on the particle in addition to fluid-mediated interactions between particles. Given an accurate and consistent set of models for the particle-scale dynamics (e.g., $\textbf{F}_p$, $\textbf{R}_g$, and $Q_p$), such a framework shows promise for enabling large-scale simulations of high-speed two-phase flows.

\subsection{Energy Transfer}
\begin{figure}[h]
\includegraphics[width=0.85\textwidth]{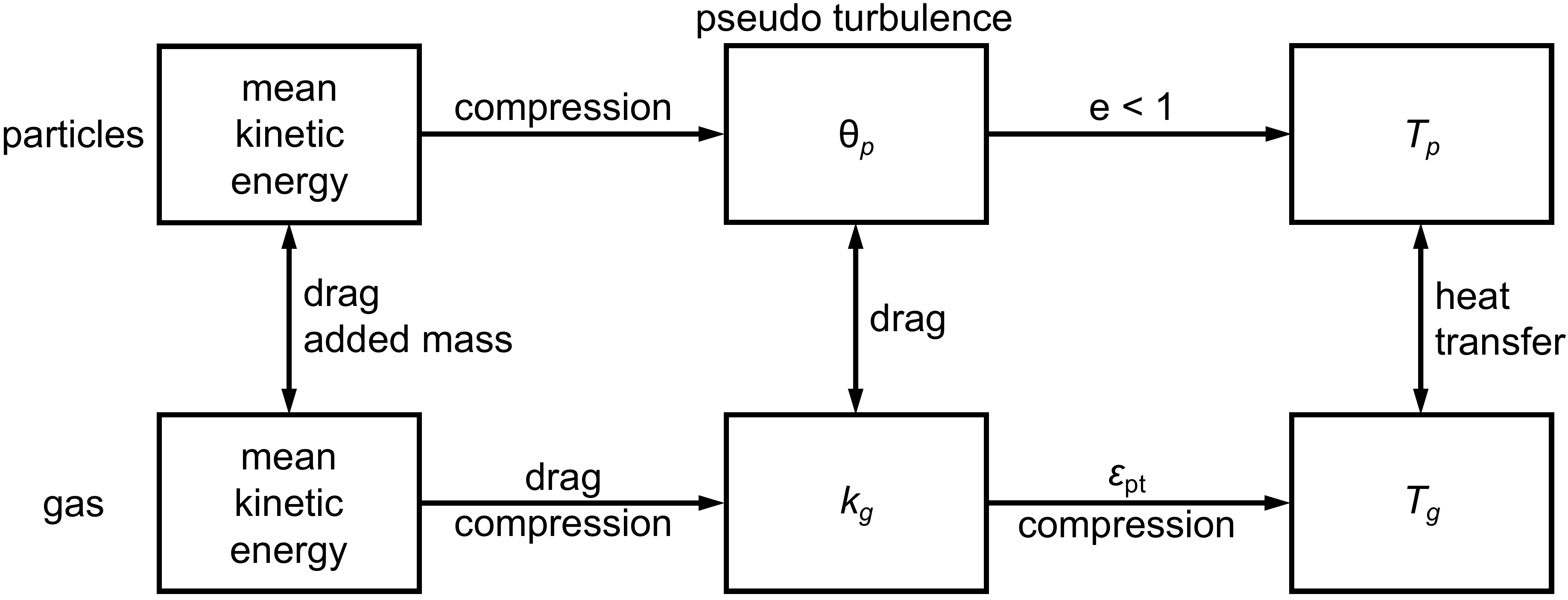}
\caption{Energy diagram for compressible gas-particle flows. In this description, the total gas-phase energy is given by $E_g=\frac{1}{2}u_g^2+k_g+e_g$ and the total particle-phase energy is $E_p=\frac{1}{2}u_p^2+\frac{3}{2}\Theta_p+e_p$, where $e_g$ and $e_p$ are the internal energies of the gas phase and particles, respectively. For a closed system, all energy flows to the lower-right corner. Courtesy of RO Fox.}
\label{fig:energy}
\end{figure}

Energy transport equations can be derived from the averaged two-phase flow equations, providing a framework for multiphase turbulence modeling \citep{fox2014}. \textbf{Figure~\ref{fig:energy}} shows how energy is transferred in compressible gas-particle flows. Particles exchange mean kinetic energy with the carrier phase through drag and added mass. This large-scale motion in the gas phase is then transferred to pseudo-turbulence, $k_g$, by the random arrangement of particles and non-linear interactions between wakes. A portion of $k_g$ generates random uncorrelated motion in the particle phase (termed granular temperature; $\Theta_p$), and the remainder is dissipated to heat via viscous dissipation, $\varepsilon_{\rm pt}$. Gas compression also contributes to turbulent kinetic energy \citep{sarkar1992pressure}, while particle compression acts as a source of granular temperature \citep{capecelatro2015on}. Granular temperature generates collisions between particles, and for inelastic particles (coefficient of restitution ${\rm e}<1$), the energy associated with collisions is dissipated to heat. 

\begin{marginnote}[]
\entry{Granular temperature}{Energy associated with random uncorrelated particle motion}
\end{marginnote}

Transport of PTKE is given by \citep{shallcross2020volume}
\begin{equation}
    \frac{\partial (1-\Phi_v)\rho_g k_g}{\partial t}+\nabla\mathbf{\cdot}\left[(1-\Phi_v)\rho_g\textbf{u}_gk_g\right]=-(1-\Phi_v)\textbf{R}_g\mathbf{:}\nabla\textbf{u}_g+\left(\mathbf{u}_p-\mathbf{u}_g\right)\mathbf{\cdot}\textbf{F}_p-(1-\Phi_v)\rho_g\varepsilon_{\rm pt}.
    \label{eq:ptke}
\end{equation}
Unlike in single-phase turbulence modeling based on the Reynolds-averaged Navier--Stokes (RANS) equations, turbulent kinetic energy has a source of production from drag (second term on the right-hand side of Equation \ref{eq:ptke}), resulting in the generation of $k_g$ even in the absence of mean shear (first term on the right-hand side of Equation \ref{eq:ptke}).
Dissipation of fluid-phase turbulence via drag takes place at length scales on the order of $d_p$. With the aide of PR-DNS, closure models appearing in the multiphase RANS equations can be formulated following the energy-flow diagram shown in \textbf{Figure~\ref{fig:energy}}.

\section{MULTI-PARTICLE INTERACTIONS}
Gas-particle interactions in flows containing many particles are discussed in this section. 
Insights gleaned from PR-DNS on the drag force exerted by compressible flows on assemblies of particles are first reviewed. The effects of interphase exchange on flow-generated sound are then discussed. We then summarize existing experimental configurations that isolate shock-particle interactions. Such flows are particularly challenging to simulate numerically, as they require methods that can simultaneously capture shock structures and the disperse phase. As already discussed, subgrid-scale models under these flow conditions are far less developed compared to flows in the absence of shocks. Experiments are challenging owing to the wide range of temporal scales and reduced optical access caused by the particles.

We focus on dilute suspensions of particles in compressible jets and dense suspensions of particles in shock tubes.
Such experiments are useful for validating numerical models and understanding the intricate dynamics shared by a broader class of compressible gas-particle flows.

\subsection{Drag at Finite Mach Number and Volume Fraction}
The drag force in systems containing many interacting finite size particles differs greatly from that of an isolated particle. This is well documented for incompressible flows \citep[e.g., see][and references therein]{tenneti2014particle}. \cite{mavcak2021regimes} developed regime maps for subsonic flow in dense gas-particle systems. Using theoretical arguments, they showed that the drag forces arising in assemblies of particles result in compressible effects at Mach numbers well below the typical incompressible criterion used for an isolated sphere (i.e., $\Mac<0.3$). With recent progress in numerical methods and growing computational resources, PR-DNS are starting to provide quantitative measures of the drag force exerted by compressible flows in dense suspensions. The green shaded region in \textbf{Figure~\ref{fig:drag}} corresponds to expected values for the (steady) drag coefficient at finite volume fraction and Mach number in the continuum regime.

PR-DNS of shock waves passing through random distributions of spherical particles reveal significant particle-to-particle variation in drag \citep{das2018strategies,mehta2019effect,osnes2021performance,osnes2022}. Particles quickly dissipate energy from the shock, causing the peak unsteady drag force to decrease with downstream distance. In inviscid flows, the formation of shocklets and bow shocks around the particles results in non-zero drag long after the shock passes. Interestingly, it has been shown that the mean drag force acting on the suspension closely matches the force acting on an isolated particle \citep{mehta2019effect}. In contrast, the mean drag force on a suspension of particles exerted by a viscous gas increases with increasing volume fraction, with values of $C_D$ significantly higher in particle clouds compared to values reported in single-particle studies \citep{osnes2019computational}.

Most PR-DNS of compressible gas-particle flows consider shock-particle interactions, which introduce challenges in developing drag correlations due to the lack of statistical stationarity and homogeneity. \citet{khalloufi2023drag} reported the first simulations of homogeneous compressible flows past random arrays of particles ranging from subsonic to supersonic free-stream Mach numbers.
The magnitude of the drag force was found to increase with volume fraction, consistent with findings from incompressible flows. Increasing the volume fraction was also found to reduce the critical Mach number that demarcates the transonic regime. 
\citet{Osenes2023drag} expanded the data set of \citet{khalloufi2023drag} to a wider range of Reynolds numbers and proposed models for quasi-steady drag, quasi-steady drag variation, and transverse (lift) forces, representing the most comprehensive drag law for spheres to date. The expressions reduce to well-established force models in the zero Mach number limit and in the isolated particle limit. The corresponding drag coefficient for $\Phi_v=0.4$ and $\Mac=0.8$ is shown in \textbf{Figure~\ref{fig:drag}}. 

It is important to note that to date there has been little (if any) comparisons of the forces measured from PR-DNS with experiments. This requires carefully setup experiments to isolate the effect of drag on particle motion, and diagnostics capable of measuring gas-phase velocity in the vicinity of particles. Canonical experimental configurations of shock-particle interactions will be summarized in Sections~\ref{sec:jet} and~\ref{sec:shock-tube}.

\subsection{Particle-Turbulence-Acoustics Interactions}
Turbulent compressible flows are capable of radiating acoustic waves that in some cases can generate significant (often undesired) sound pressure levels (SPL) \citep[e.g.,][]{tam1995supersonic}. 
Interphase coupling is capable of amplifying or attenuating SPL. For example, water injection has been observed experimentally to reduce near-field sound levels from high-speed jets by 2--6 dB using 5--16\% of mass of the gas jet \citep{krothapalli2003turbulence} and as much as 12 dB near rocket engine exhausts with $\Phi_m>1$ \citep{henderson2010fifty}. 

The precise mechanisms responsible for these observed changes in SPL are complex. The acoustic analogy introduced by \citet{lighthill1952sound} is widely used to analyze sound generated from turbulent flows. \citet{crighton1969sound} extended Lighthill's acoustic analogy to quantify the effect small air bubbles have on the sound generated from turbulent flow in water. The theory predicts that sound levels increase with $\Phi_m$, and this increase is significant when $\Phi_m>1$. This is in contrast to experimental observations of reduced sound during water injection into  high-speed jets of air. 

To better understand the effects of interphase coupling on flow-generated sound, the averaged equations of motion \ref{eq:density}--\ref{eq:energy} can be rearranged to arrive at a transport equation for the gas-phase pressure, given by \citep{buchta2019sound}
\begin{equation}
    \frac{\partial p_g}{\partial t}+\mathbf{u}_g\mathbf{\cdot}\nabla p_g=-\gamma p_g\nabla\mathbf{\cdot}\mathbf{u}_g-\mathcal{D}+\frac{\gamma-1}{1-\Phi_v}\left(\mathbf{u}_p-\mathbf{u}_g\right)\mathbf{\cdot}\textbf{F}_p+\gamma p_g\frac{{\rm D}\ln{\Phi_v}}{{\rm D}t},
    \label{eq:pressure}
\end{equation}
where terms involving molecular transport effects are combined into $\mathcal{D}$. 
Compared to the pressure transport equation for single-phase flows \citep[e.g.,][]{pantano2002study}, the last two terms on the right-hand side represent new
contributions due to particles. The last term accounts for volume displacement effects through changes in $\Phi_v$, representing a $pDV$ work term due to particles entering or leaving a control volume \citep{houim2016multiphase}. The term involving $\textbf{F}_p$ accounts for work due to drag and is only active when the slip velocity between the phases is non-zero.

\begin{figure}[h]
\includegraphics[width=0.95\textwidth]{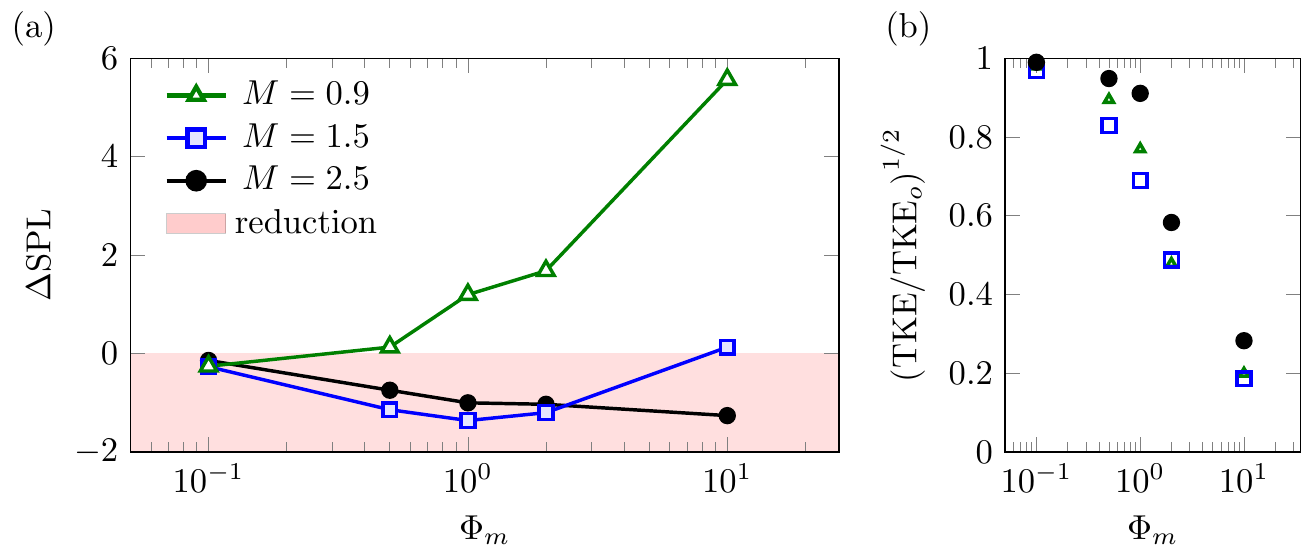}
\caption{Results from numerical simulations of compressible mixing layers seeded with particles of Stokes number ${\rm St}=1$ at different convective Mach numbers ($M$). (\textit{a}) Sound pressure level changes at a distance above the mixing layer. (\textit{b}) Reduction in turbulence levels at the centerline of the mixing layer. Reproduced with permission from \cite{buchta2019sound}.}
\label{fig:sound}
\end{figure}

\citet{buchta2019sound} performed numerical simulations of particle-laden, high-speed shear layers to quantify the effect of inertial particles on turbulence and near-field pressure intensity. Interphase coupling was observed to have a broadband effect on the turbulence and pressure
fields, reducing the turbulence by over 70\% for $\Phi_m=10$ (see \textbf{Figure~\ref{fig:sound}}). Volume displacement and drag coupling were found to have competing effects on the local pressure fluctuations. For subsonic flow, the sound level increases with particle loading, consistent with low-Mach number multiphase aeroacoustic theory. As shown in \textbf{Figure~\ref{fig:sound}}, the SPL increased despite a marked decrease in turbulent kinetic energy (TKE). The increase in SPL was found to be due to changes in gas compression by the particles due to volume displacement and drag. In contrast, the sound levels were found to decrease with increasing mass loading in supersonic flows, largely due to the substantial decrease in TKE. While the field of aeroacoustics has received much attention in the past half-century, multiphase aeroacoustics remains largely unexplored. The introduction of a disperse phase may provide alternative sound-reducing mechanisms compared to traditional passive and active control methods.

\subsection{Shock-Particle Interactions in Jets}\label{sec:jet}
\begin{figure}[h]
\includegraphics[width=0.9\textwidth]{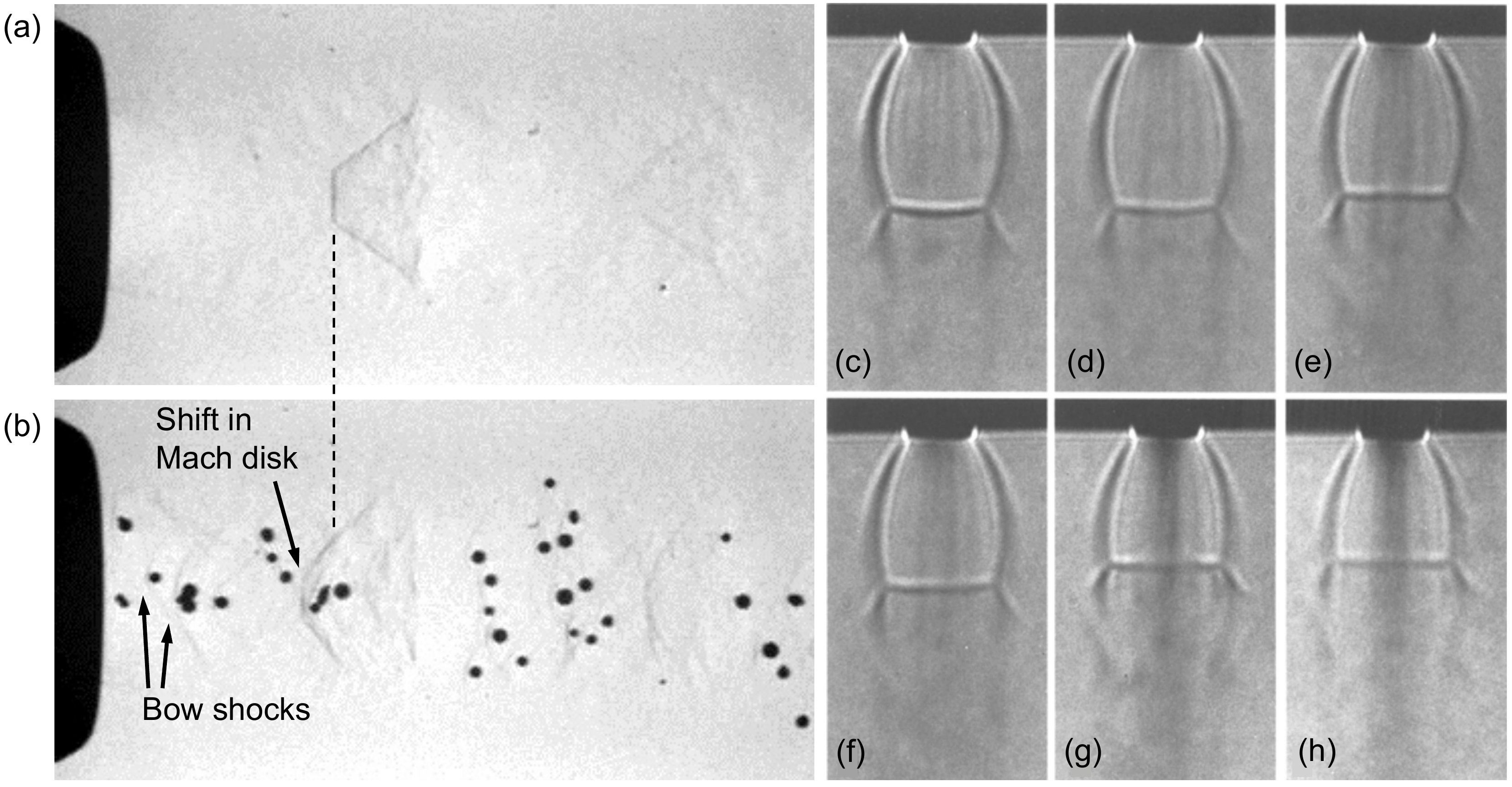}
\caption{Experimental visualizations of underexpanded jets. (\textit{a}) Schlieren of an unladen jet with ${\rm NPR}=4.76$ and $D_e=2$ mm. (\textit{b}) Same jet laden with $100$ $\upmu$m particles. (\textit{c})--(\textit{f}): Shadowgraph of an underexpanded jet with ${\rm NPR}=29.8$, $D_e=3$ mm, and $d_p=45$ $\upmu$m for different mass loadings: (\textit{c}) $\Phi_m=0$; (\textit{d}) $\Phi_m=0.11$; (\textit{e}) $\Phi_m=0.24$; (\textit{f}) $\Phi_m=0.35$; (\textit{g}) $\Phi_m=0.64$; and (\textit{h}) $\Phi_m=1.07$. Panels \textit{a}--\textit{b} courtesy of JS Rubio. Panels \textit{c}--\textit{f} adapted from \cite{sommerfeld1994structure}.}
\label{fig:jet}
\end{figure}

Experimental studies on particle-laden compressible jets date back to the 1960s \citep{bailey1961gas,hoglund1962recent,marble1963nozzle,lewis1964normal,jarvinen1967underexpanded}. As shown in \textbf{Figure~\ref{fig:jet}}, particles tend to modify the shock structures within the jet. In an unladen underexpanded jet, the location of the Mach disk is a function of the ratio of the total (tank) pressure to the ambient pressure, or nozzle pressure ratio (NPR).
A common correlation for the location of the Mach disk is $L_{\rm MD}/D_e=\sqrt{{\rm NPR}/2.4}$, where $D_e$ is the nozzle diameter \citep{crist1966study}. 
\citet{lewis1964normal} developed a mass-loading-dependent correction factor $f$ based on an empirical fit to experimental data of micron sized alumina particles in high-speed jets, given by $f(\Phi_m,\Mac_e)=(1+0.197\Mac_e^{1.45}\Phi_m^{0.65})^{-1}$, where $\Mac_e$ is the Mach number at the nozzle exit. For a supersonic jet with $\Mac_e=3$, this predicts a 30\% shift in $L_{\rm MD}$ when $\Phi_m\approx 0.3$, consistent with earlier experimental findings. Even larger changes would be predicted at higher exit Mach numbers. Thus, even for relatively dilute systems, particles are capable of significantly modifying the carrier phase, and the coupling becomes more pronounced at higher Mach numbers. Meanwhile, the mechanisms causing these changes remain unclear.

More recently, \citet{sommerfeld1994structure} performed experiments and numerical simulations of particle-laden underexpanded jets at ${\rm NPR}=30$. \textbf{Figure~\ref{fig:jet}} shows shadowgraph images of the jet with increasing values of $\Phi_m$. The shift in Mach disk was attributed to strong momentum coupling between the phases due to the large relative velocity between the gas and particles. Their experiments significantly overpredict the shift in $L_{\rm MD}$ compared to the correlation of \citet{lewis1964normal}, with smaller particles resulting in a larger shift. This was attributed to smaller particles influencing a larger portion of the jet due to their increased spreading rate. Eulerian--Lagrangian simulations were capable of predicting a shift in the Mach disk location, but not to the same degree as was seen in the experiments. 

Further work is needed to better understand the exact mechanisms responsible for the modifications to the shock structures, and expose limitations of numerical models.  In general, the best way to improve fundamental understanding on shock-particle interactions is numerical simulation concurrent with advanced experimental methods (see sidebar titled Experimental Diagnostics for High-Speed Multiphase Flow).

\begin{textbox}[h]\section{EXPERIMENTAL DIAGNOSTICS FOR HIGH-SPEED MULTIPHASE FLOW}
The last two decades have seen rapid advances in high-speed imaging technology. Digital cameras with high spatial resolution, MHz repetition-rates, and long record times are becoming common, allowing image-based particle tracking in supersonic flows. Additionally, the advent of high-repetition-rate, high-power illumination sources such as pulse-burst lasers, has enabled time-resolved particle image velocimetry (TR-PIV) to characterize the gas phase velocity \citep{beresh2021time}. In such experiments, care must be taken to separate the typically slower particles of interest from the PIV seed \citep{demauro2017unsteady}. With the continuous development of sophisticated algorithms, digital in-line holography (DIH) can now resolve 3D particle velocity and size distributions in shock tubes \citep{chen2018galinstan} and in underexpanded jets \citep{buchmann2012ultra}. Researchers have resorted to X-ray techniques in dense multiphase flows opaque to visible light. For instance, single-shot, flash X-ray has measured particle volume fraction in dense clouds \citep{wagner2015flash}, whereas time-resolved proton radiography has successfully tracked explosively dispersed particle beds \citep{hughes2021proton}. Algorithms combining 3D tracking with X-ray are promising for optically opaque media but are only in their infancy with demonstrations limited to creeping flow \citep{makiharju2022tomographic}.
\end{textbox}

\subsection{Particle Curtain Interactions in a Shock Tube}\label{sec:shock-tube}
\begin{figure}[h]
\includegraphics[width=0.7\textwidth]{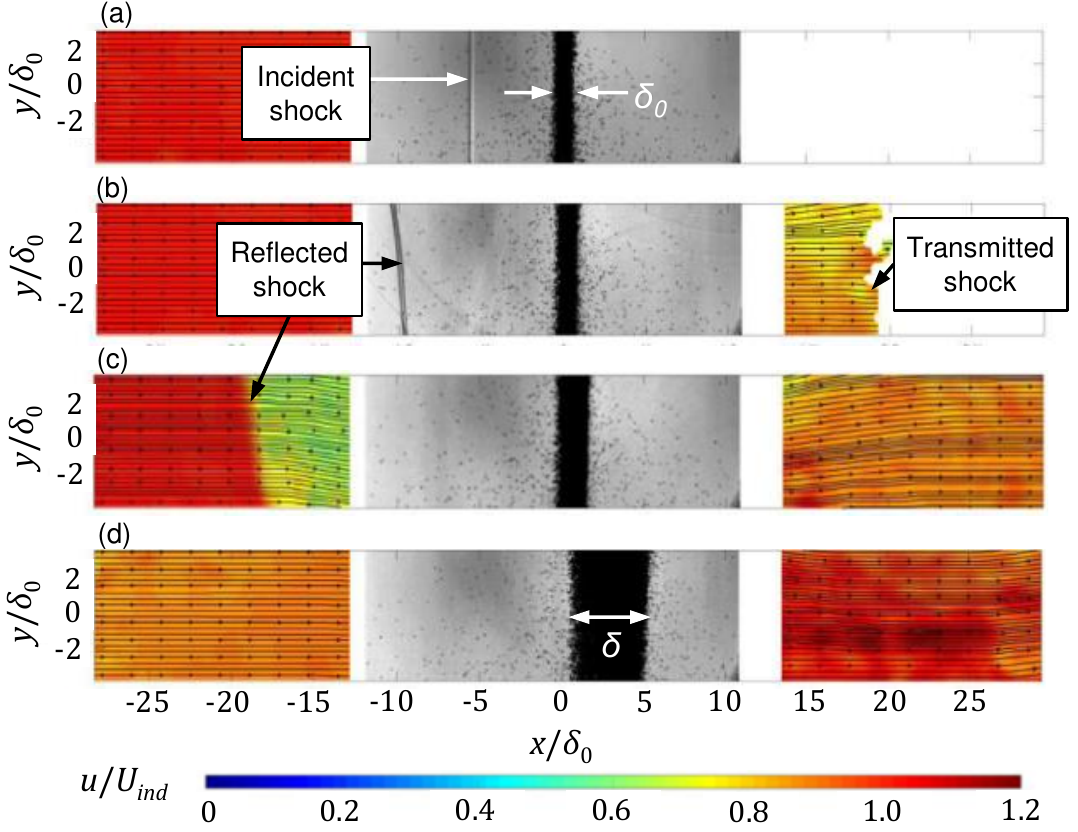}
\caption{Interaction of a Mach number $\Mac_s=1.22$ normal shock with a particle curtain having initial volume fraction $\Phi_v=0.09$ and initial thickness $\delta_0$ = 3.0 mm. Streamwise velocity $u$ contours normalized by post-shock incident shock velocity $U_{\rm ind}$. Normalized time $t^*$ given by Equation \ref{eq:vol_frac_time}. At $t^*$ = (\textit{a}) $-0.018$ ($-40$ $\upmu$s), (\textit{b}) $0.067$ ($147$ $\upmu$s), (\textit{c}) $0.128$ ($280$ $\upmu$s), and (\textit{d}) $0.287$ ($627$ $\upmu$s). Adapted from \cite{demauro2017unsteady}.}
\label{fig:DeMauro}
\end{figure}

%
%
Several experimental campaigns have sought to study dense gas-solid flows using shock tubes. The pioneering work of \cite{rogue1998experimental} studied shock-particle interactions at nearly packed volume fractions of $\Phi_v\approx0.6$ by placing a bed of spheres on a thin diaphragm in a vertical shock tube. 
Later, researchers targeted shock-particle interactions in dense regimes using gravity-fed particle curtains \citep[e.g.,][]{wagner2012multiphase,theofanous2016dynamics,demauro2017unsteady,theofanous2018shock,demauro2019improved,daniel2022shock}. An example is shown in \textbf{Figure~\ref{fig:DeMauro}}, which includes schlieren imaging of the particle curtain and visualized streamwise gas-phase velocity measured with TR-PIV.
\begin{marginnote}[]
\entry{TR-PIV}{An optical experimental measurement technique that captures temporal and spatial velocity information}
\end{marginnote}
The incident shock (\textbf{Figure~\ref{fig:DeMauro}a}) is transmitted and reflected by the particles (\textbf{Figure~\ref{fig:DeMauro}\textit{b}}) resulting in a pressure gradient across the curtain. This pressure differential is the largest contributor to unsteady drag of the particle curtain \citep{demauro2017unsteady}. At later times, the curtain spreads (\textbf{Figure~\ref{fig:DeMauro}\textit{d}}). 
Regions of high velocity are observed within the curtain, which are likely correlated to changes in porosity that act effectively as a nozzle to accelerate the local flow (e.g., see \textbf{Figure~\ref{fig:shock-particle}}). 

\begin{figure}[h]
\includegraphics[width=.9\textwidth]{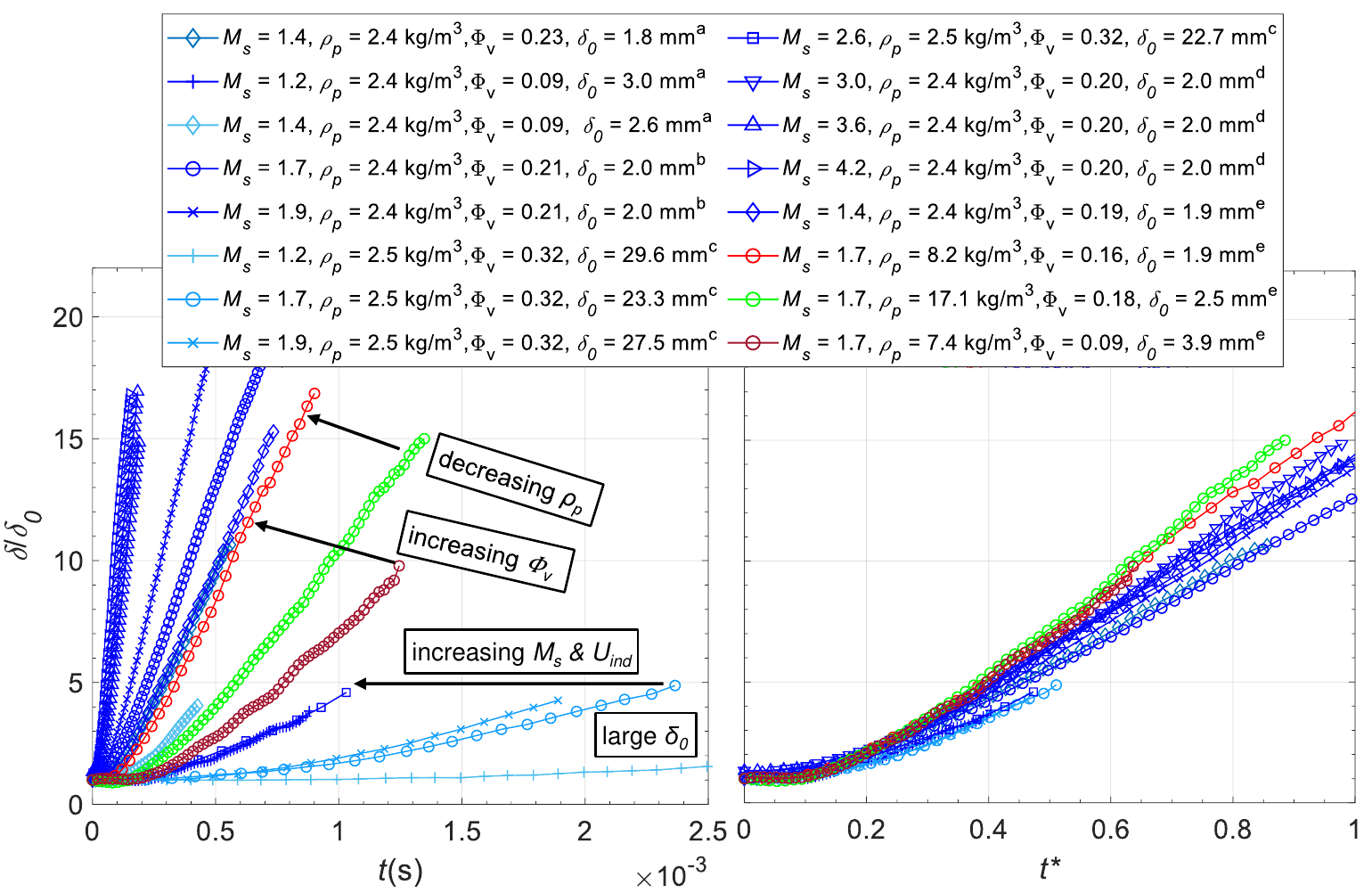}
\caption{Comparison of particle curtain spread data. (\textit{a}) Non-dimensional spread $\delta$/$\delta_0$ versus time, and (\textit{b}) $\delta$/$\delta_0$ versus $t^*$ (Equation \ref{eq:vol_frac_time}). Glass particle trajectories are shown in blue, tungsten in green, and steel in red. Adapted from \citet{wagner2023shock}.}
\label{fig:Curtain_Scaling}
\end{figure}

The multitude of particle curtain experiments performed to date span a wide range of $\Mac_s$, $\rho_p$, $\Phi_v$, and initial curtain thickness $\delta_0$. The dependence of the non-dimensional curtain spread $\delta/\delta_0$ on these parameters is shown in \textbf{Figure~\ref{fig:Curtain_Scaling}\textit{a}}. Not surprisingly, particles spread faster as $\Mac_s$ increases and $\rho_p$ decreases. The experiments of \cite{theofanous2016dynamics} with larger $\delta_0$ show a markedly lower spreading rate. Also evident is the faster spread of the curtain with increasing $\Phi_v$. To explain the spread, \cite{theofanous2016dynamics} normalized time as a function of the theoretical reflected shock pressure were the curtain a solid wall. Expanding upon this, \citet{demauro2019improved} and \citet{daniel2022shock} used a force balance approach to suggest the following scaling relationships:
\begin{equation}
\frac{x}{\delta_0}= \left({\frac{\sqrt{\Delta P}t}{\sqrt{\rho_p}\delta_0}}\right)^2,
\label{eq:pressure_time}
\end{equation}
and
\begin{equation}
\frac{x}{\delta_0}{\displaystyle \propto}~\left(\Phi_v^{0.25}\sqrt{\frac{\rho_0}{\rho_p}}\frac{U_{\rm ind}t}{\delta_0}\right)^2={t^*}^{2},
\label{eq:vol_frac_time}
\end{equation}
where $\Delta P$ is the pressure difference across the curtain, $\rho_0$ is the initial gas density, and $U_{\rm ind}$ is the theoretical velocity induced by the incident shock. \citet{daniel2022shock} demonstrated tight collapse of the particle curtain spread using Equation~\ref{eq:pressure_time} and the pressures measured across the curtain over a range of volume fractions. 
Alternatively, the time scaling in Equation~\ref{eq:vol_frac_time} is derived by treating the particle curtain as a porous screen and incorporates only parameters known a priori. As shown in \textbf{Figure~\ref{fig:Curtain_Scaling}\textit{b}}, this scaling collapses the curtain spread over nearly an order of magnitude range in $U_{\rm ind}$ ($110-1170$ m/s), $\rho_p$ ($2.4-17.1$ kg/m$^3$) and $\delta_0$ ($1.8-33.5$ mm). 
Notably, the particle diameter is not included in the scaling, although it also varies by an order of magnitude ($0.1-1$ mm). This suggests that the unsteady dynamics of shock-induced dispersal are dominated by properties within the curtain as it expands, and that $\delta_0$ is a more important length scale than $d_p$ at late times.

Despite progress, several questions on shock-particle curtain interactions remain. The parameter space should be expanded to include high density particles with stronger shocks. Moreover, as the particle cloud expands and becomes more dilute, the dense scaling will necessarily fail. For instance, \cite{theofanous2018shock} report good agreement with standard drag laws when the initial volume fraction within the curtain is less than 1\%. Experiments with initial volume fraction of a few percent would expose this dense-to-dilute transition. Additionally, all experiments have been performed with monodisperse particles. Experiments and numerical simulations of bidisperse and polydisperse suspensions would provide insight into shock-induced size segregation (e.g., \textbf{Figure~\ref{fig:shock-particle}}).

Multiphase shock tubes represent one of the few experimental configurations commonly used for validating compressible gas-particle flow models. Over the past decade, several numerical studies have reported good agreement in the curtain spreading rate and location of reflected and transmitted shocks, but demonstrate that the results depend strongly on the treatment of the unclosed terms \citep[e.g.,][]{ling2012interaction,houim2016multiphase,shallcross2018parametric}. Discrepancies between simulations and experiments can be traced to experimental uncertainty in the initial particle distribution and boundary layers at the walls of the shock tube. Improvements to quasi-steady and unsteady forces and particle collisions are needed to reduce errors in long-term predictions of the downstream particle front position \citep{nili2021prioritizing}. There is a need for experimental measurements with improved uncertainty quantification and further parametric studies to support model development.

Even on modern computers, particle-resolved simulations running long enough to observe spreading of the curtain ($t^*\gtrapprox0.2$) remain out of reach. To ensure numerical stability in an explicit discretization of the compressible Navier--Stokes equations, the simulation time step must be $\Delta t<\Delta x/\max{|\mathbf{u}_g+c|}$, where the grid spacing used in PR-DNS is $\Delta x \ll d_p$. As an example, consider the simulation shown in \textbf{Figure~\ref{fig:shock-particle}}. The parameters closely match the multiphase shock tube experiment of \citet{wagner2012multiphase}. Uniform grid spacing is employed with 40 points across the diameter of the smallest particles, resulting in $\Delta t\approx10^{-9}$ s, about $10^{6}$ smaller than the time span in \textbf{Figure~\ref{fig:Curtain_Scaling}}. Recent advances in immersed boundary methods using adaptive mesh refinement \citep{mehta2022particle} show promise in addressing some of these issues. Combining such techniques with multirate time integration \citep[e.g.,][]{mikida2019multi} or all-Mach number solvers \citep[e.g.,][]{kuhn2021all} could aid in reconciling the timescale discrepancy.

\begin{summary}[SUMMARY POINTS]
\begin{enumerate}
\item Compared to incompressible flows, typical particle-laden compressible flows span a much wider range of scales. In addition, unsteady forces that are typically negligible in low-speed gas-particle flows (e.g., added mass and Basset history) can affect flows with large gas-phase acceleration, especially during shock-particle interactions. Validated models for such interactions exist when the post-shock Mach number is subcritical.
\item For a given free-stream Mach number, gas-phase compressibility becomes increasingly more important as the particle volume fraction increases.
\item Existing drag laws for particles at finite Mach number have surprising history from 18th- and 19th-century cannon firings, which only loosely matches the needed parameter space. 
\item Compressible two-fluid models are known to become ill-posed due to lack of hyperbolicity when two-way coupling is accounted for. After nearly 40 years of attempts to remedy this, a fully hyperbolic two-fluid model was recently formulated \citep{fox2019kinetic,fox2020hyperbolic}.
\item Particle-resolved simulations are shedding new light on drag and turbulence when shock waves interact with assemblies of particles.
\item Advances in high-speed measurement diagnostics and novel experimental configurations over a wide range of volume fractions have provided insight into unsteady forces associated with shock-particle interactions, motivating a multitude of modeling efforts. 
\end{enumerate}
\end{summary}

\begin{issues}[FUTURE ISSUES]
\begin{enumerate}
\item Improved numerical methods are needed to properly account for particles in the vicinity of shocks when the grid spacing is larger than the particle diameter.
\item Separate (but compatible) models for intrinsic turbulence (turbulence that follows the classical energy cascade) and pseudo-turbulence (turbulence generated by particles at small scales) are needed.
\item  Careful experiments resolving particle motion and the surrounding fluid are needed to validate numerical simulations. To resolve the flow-field, such experiments will likely use larger particles, challenging current DNS capabilities.
\item Particle size segregation during shock-driven expansion, such as in explosive dispersal, is poorly understood. Numerical simulations and experiments of shocks interacting with bidisperse and polydisperse mixtures and nonspherical particles will provide important insights.
\item Multiphase instabilities observed in strongly accelerating flows \citep[e.g.][]{rodriguez2013solid,mcfarland2016computational,frost2018heterogeneous,osnes2018numerical} remain poorly understood and need to be unraveled.
\item Flow through a collection of particles results in a distribution of drag forces with significant particle-to-particle variation. Various models have recently been proposed to capture drag force variation in incompressible \citep{akiki2017pairwise,esteghamatian2018stochastic,lattanzi2022stochastic} and compressible \citep{Osenes2023drag} flows. Future research should incorporate and assess the importance of such models in coarse-grained simulations.
\item Validated models for interphase heat and mass transfer that account for reacting particles at finite $\Mac$ and $\Phi_v$ are needed. Recent progress in this area can be found in \citet{ling2016inter,houim2016multiphase,das2018metamodels}.
\item At hypersonic speeds, temperature effects become important and ionization of the carrier phase might need to be accounted for. Recent advances have been made in heat flux and drag modeling under these conditions \citep[e.g.,][]{singh2016heat,singh2017aerothermodynamic}, including the transition from rarefied to continuum flows in dense particle suspensions \citep{vijayan2022kinetic}. However, detailed experiments are required to validate and identify the limitations of these models.
\end{enumerate}
\end{issues}

\section*{DISCLOSURE STATEMENT}
The authors are not aware of any affiliations, memberships, funding, or financial holdings that might be perceived as affecting the objectivity of this review. 

\section*{ACKNOWLEDGMENTS}
J.C. acknowledges the support from the National Aeronautics and Space Administration (grant no. 80NSSC20K1868 and 80NSSC20K0295). He is also grateful for contributions from current and former students and postdocs, including Dr. Gregory Shallcross, Dr. Mehdi Khalloufi, Meet Patel, and Archana Sridhar. J.W. is grateful for support from the Laboratory Directed Research and Development Program (LDRD). He thanks Steven Beresh, Sean Kearney, Edward DeMauro, Kyle Daniel, and Daniel Guildenbecher for critical input and insightful discussions. 

Sandia National Laboratories is a multi-mission laboratory managed and operated by National Technology and Engineering Solutions of Sandia, LLC, a wholly owned subsidiary of Honeywell International Inc., for the U.S. Department of Energy’s National Nuclear Security Administration under contract DE-NA0003525. This paper describes objective technical results and analysis. Any subjective views or opinions that might be expressed in the paper do not necessarily represent the views of the U.S. Department of Energy or the United States Government.

\bibliographystyle{ar-style1}
\bibliography{main}

\end{document}